# DSA, AIA, and LLMs: Approaches to conceptualizing and auditing moderation in LLM-based chatbots across languages and interfaces in the electoral contexts

Natalia Stanusch, Raziye Buse Çetin, Salvatore Romano, Miazia Schueler, Meret Baumgartner, Bastian August and Alexandra Roșca

## Abstract

The integration of Large Language Models (LLMs) into chatbot-like search engines poses new challenges for governing, assessing, and scrutinizing the content output by these online entities, especially in light of the Digital Service Act (DSA). In what follows, we first survey the regulation landscape in which we can situate LLM-based chatbots and the notion of moderation. Second, we outline the methodological approaches to our study: a mixed-methods audit across chatbots, languages, and elections. We investigated Copilot, ChatGPT, and Gemini across ten languages in the context of the 2024 European Parliamentary Election and the 2024 US Presidential Election. Despite the uncertainty in regulatory frameworks, we propose a set of solutions on how to situate, study, and evaluate chatbot moderation.

**Keywords**: Copilot, LLMs, moderation, elections, chatbots.

# Introduction

From OpenAI's introduction of ChatGPT to Microsoft's inclusion of Copilot into its search engine Bing, chatbots with integrated Large Language Models (LLMs) are alleged to aid users in finding relevant information (OpenAI [A]) and accessing online content (Microsoft, Oct 1, 2025). Aside from gaining popularity – in 2024, over 200 million active users turned to ChatGPT weekly (Reuters, Aug 29, 2024) – initial changes in website traffic indicate that more and more users turn to LLM chatbots than to 'traditional' search engines (Bianchi and Angulo, Nov 20, 2024). LLMs are trained on unprecedented amounts of data scraped from internet forums, articles, and more (Koebler, Jan 29, 2025), making use of the increase in computing power and deep neural network logics. By generating text strings grounded in patterns and correlations among words, LLMs accurately mimic human language and perform a variety of different tasks (IBM., Nov 2, 2023).

But with the advancement of LLMs and their adaptation into chatbots, concerns over the amplification of biases and questionable ethical standards arise (Ranjan, Gupta, and Singh, 2024; Gibney, 2024; UNESCO 2024; Navigli, Conia, and Ross 2023). Due to the probabilistic nature of chatbots, their outputs might 'sound' accurate yet contain misleading or critical errors. Indeed, LLM-based chatbots, in principle based on predicting the most likely correlation of words, do not have innate ways of fact-checking information and lack transparency regarding information selection (Zewe 2024; Augenstein et al., 2023). Consequently, urgent questions arise on who should take responsibility for faulty, misleading, or unsafe content these chatbots output – and who should step in to prevent these models from causing harm.

In its relatively new history of public use, LLM-based chatbots have already proven unreliable and disruptive to public discourse, ranging from spreading hate speech and fake information (Victor, March 24, 2026) to inappropriate user interactions (Roose, Feb 16, 2023). Existing studies have already demonstrated how LLMs produce factual errors and may thus spread false information (Angwin et al., 2024; Yao et al., 2024; Zhang et al., 2023; Romano et al., 2023). LLM-based chatbots were also recently scrutinized for the lack of safeguards, outputting misinformation and disinformation, as well as conspiracy theories on topics such as the Russian invasion of Ukraine, climate change, and the Holocaust (Angwin et al., 2024; Kivi, 2024; Simon et al., 2024; Urman and Makhortykh, 2024; Kuznetsova et al., 2023). LLM-based chatbots also inaccurately summarized, quoted, and attributed information reported by news outlets, such as the BBC (Elliott, Feb 2025).

In light of these alarming reports, the companies behind LLM-based chatbots have taken different approaches to deploy additional safeguards. However, the risk management of LLMs and their downstream applications is complex, and the regulatory approaches are evolving to address these challenges. For instance, when ChatGPT was released in November 2022, the Digital Services Act (the DSA), an EU regulation geared towards regulating online platforms broadly, had already been adopted in the EU.[1] The draft AI Act (the AIA) was not equipped to address the risks of General Purpose AI Systems (GPAIs; which include LLMs) because its risk-based approach was heavily dependent on the use case of the AI system. With the fast deployment of ChatGPT, the AIA's scope and risk management framework were reframed to incorporate provisions aiming to regulate GPAI models.[2] As such, the AIA entered into force on August 1, 2024. Yet it remains unclear how different regulatory approaches will apply in practice to LLM-based chatbots and whether they can effectively address their information-related risks.

---

[1] The Digital Services Act was published in the Official Journal on October 27, 2022 and entered into force on November 16, 2022. The full application of the DSA provisions came into force on February 17, 2024.

A prime testing ground to verify how these challenges of content generation and retrieval are handled is the topic of elections. The topics surrounding elections are particularly relevant given the requirement to implement the Digital Services Act (DSA), as election integrity is among the systemic risks that Very Large Search Engines (VLOSEs) are explicitly called to mitigate. One example of such a VLOSE with LLM functionality is Bing's Copilot. Thus, this chapter focuses on answering the following research questions:

How can we conceptualize LLM-based chatbots in relation to existing regulatory frameworks and related fields, such as online platform moderation? What methodologies, approaches, and strategies can we use to independently evaluate moderation in LLM-based chatbots? How can we define and measure the current application of moderation on LLM-based chatbots such as Copilot, ChatGPT, and Gemini, by taking as the case study the 2024 European Parliamentary election and the 2024 US presidential election?

To answer these questions, this chapter includes three interlacing sections. First, we survey the landscape of moderation, regulation, and policy in which we can situate LLM-based chatbots. LLM-based chatbots are discussed in relation to former platform moderation practices and critiques. We also acknowledge the regulatory difficulty in considering chatbots as platforms in the scope of the most recent EU regulatory framework, yet we suggest that certain notion of 'moderation' coming from platform studies proves useful when applied to LLM-based chatbots. Second, we outline the methodological approaches required to conduct our empirical study and its roots in the tradition of prompting as a method. Taking from the tradition of platform studies, we adapt the notion of 'active' and 'passive' moderation to LLM-based chatbots. We then discuss each methodological approach we implemented in our study, which consisted of auditing LLM-based chatbots by introducing a (i) cross-platform comparison that takes into account the (ii) cross-language and (iii) cross-election analysis. We then discussed the results of this study, which investigated Microsoft's Copilot, OpenAI's ChatGPT, and Google's Gemini, prompting them in ten different languages, in the context of both the 2024 European Parliamentary election and the 2024 US presidential election. Third, we argue that given the major inconsistencies in moderation across languages, coherent regulatory and scrutiny mechanisms are necessary. We highlight two key risk areas that we encountered in our studies: using the chatbots to produce propaganda 'as a service' and being exposed to misinformation 'as a default.' Given the contemporary nature of the study object, we summarize a (very recent) history of regulatory frameworks introduced and how they can (and cannot) adequately regulate LLMs in light of our findings.

## Platform Regulation and the Moderation Landscape

Evaluating the safeguards of LLMs' outputs is deeply intertwined with the history of online content moderation, primarily as understood in the platform context. From the platform moderation perspective, the practices of moderation can be framed as either a denial of access to information or a prevention of harmful content by means of reducing its accessibility or visibility. Moderation of online content constitutes a balancing act between the platform's

---

[2] Following amendments by the French and Czech Presidency, the Council adopted its position for a trilogue in November 2022 and the general approach was agreed upon on December 6, 2022.

"openness and control" (Poell et al., 2021), which internalizes the conflict between the two colliding yet central values of free speech and community protection (Gillespie, 2018). Similar practices have already been implemented in the history of the web, such as in the case of accessing specific websites and banning IP addresses (Deibert, 2008). With the rise of social media platforms, content moderation has become a necessary practice that nonetheless retains many blind spots from theoretical and empirical research perspectives (Gillespie, 2018; Gorwa, 2024; Ma et al., 2023; Tarvin and Stanfill, 2022). Similar challenges arise with the development of LLM-based chatbots.

According to the DSA, the main regulatory framework addressing online speech and content moderation in the EU, content moderation

> means the activities, whether automated or not, undertaken by providers of intermediary services, that are aimed, in particular, at detecting, identifying and addressing illegal content or information incompatible with their terms and conditions, provided by recipients of the service, including measures taken that affect the availability, visibility, and accessibility of that illegal content or that information, such as demotion, demonetisation, disabling of access to, or removal thereof, or that affect the ability of the recipients of the service to provide that information, such as the termination or suspension of a recipient's account.[3]

However, the DSA applies to user-generated content and is geared towards intermediary service providers. While the AIA is supposed to cover AI systems, including LLM-based chatbots, it does not address content moderation, freedom of expression, or information-related risks and harms (Botero Arcila, 2023). From this perspective, extending content moderation discussions to LLM-based chatbots is not self-evident. However, LLMs are increasingly integrated into intermediary services such as social media platforms and search engines. Although there is an increasing attention to the gray area on the intersection of content moderation and LLMs (Rajput, Shah, Neema, 2023; Kuai et al., 2024), the current discussion deserves more investment in conceptualization and methodological innovation for better scrutiny and effective regulatory approaches.

The datasets and 'moderation' decisions of companies that develop and release LLM-based chatbots are not transparent or accessible for external research and scrutiny. Since we do not know enough empirically about how moderation in LLM-based chatbots works, we need more insights into the actual outputs of these models to make informed policy decisions. Our previous studies show that chatbots can be used to create "propaganda as a service" by actively suggesting the production of disinformation (Romano et al., 2024) as well as spreading "misinformation by default" (Romano et al., 2023) by outputting factual errors on election-related queries. Even if some companies release transparency reports on the use of their chatbots by malicious actors to produce propaganda (OpenAI [B], 2024), we are lacking ways of independently verifying those reported claims. Thus, we actively try to create methods to scrutinize LLM-based chatbots in terms of the accuracy of the information retrieval and the loopholes for harmful content creation.

Given that democratic processes such as elections are topics susceptible to potential misinformation and manipulation, we focus on two cases of significant elections: the 2024 European Parliamentary election and the 2024 US Presidential election. In fact, the integration of chatbots, especially into search engines around 2024, coincided with a critical regulatory period:

---

[3] European Parliament and Council of the European Union. (2022, October 27). *Regulation (EU) 2022/2065 of the European Parliament and of the Council of 19 October 2022 on a Single Market for Digital Services and amending Directive 2000/31/EC (Digital Services Act)*. Official Journal of the European Union, L 277, 1–102. https://eur-lex.europa.eu/legal-content/EN/TXT/?uri=CELEX%3A32022R2065.

the DSA had just come into effect, while the AIA was still under negotiation. This period also preceded key electoral events in the EU, such as the European Parliamentary elections in June 2024, amplifying concerns about the impact of LLMs on electoral integrity. Furthermore, the occurrence of elections constitutes a significant testbed for implementing the DSA. In the DSA framework, the European Commission designates platforms and search engines with over 45 million users in the EU as Very Large Online Platforms (VLOPs) and Very Large Online Search Engines (VLOSEs). This classification subjects these entities to stringent obligations, including identifying, assessing, and mitigating "systemic risks," including risks related to civic discourse and electoral processes. Similarly, the European Commission issued Guidelines under the DSA for the Mitigation of Systemic Risks for Election in April 2024.[4] These guidelines identified both the creation and dissemination of generative AI content as sources of systemic risk, requiring VLOPs and VLOSEs to implement risk assessment and mitigation measures.

## 'New' Concepts of LLM-based Chatbot Moderation

### Emerging methodologies: prompting as research

To assess chatbot moderation, we are building our methodology on top of Rogers' Search as Research Methodology (Rogers, 2013). This approach was previously used to assess the favoring of content and sources when studying large search engines and uncovering the production of biases through source hierarchies. Using this method, we perform an algorithmic audit of the LLM-based chatbots' outputs: not of preferred sources, but rather of the occurrence of moderation that prevents the non-deterministic chatbot from generating a response (Brown et al., 2021; Rogers, 2013). Therefore, we provide the chatbot with prompts rather than search queries to assess which inputs/outputs trigger the moderation. We refer to this approach as *prompting as research*, a term previously used in AI Forensics' reporting (Romano et al., 2023; 2024) and a practice also discussed by Gillespie (2024). Prompting as research examines the politics of visibility through the diversity of the generated outputs to assess model biases and systematically assess the normative identities and narratives that are reproduced.

By testing various prompts, we attempt to achieve an algorithmic baseline of how moderation in LLM-based chatbots is implemented in relation to election content. As a general practice in researching machine learning through the comparison of inputs and outputs, we incorporate elements of *counterfactual analysis* to identify variables, such as election-related keywords and sets of words, within prompts that might trigger (or not) moderation (Cheng et al., 2024; Mishra et al., 2024). This approach assesses not only the application of chatbot moderation but also the possible causality of election-related keywords moderation to reach a level of explainable causality in chatbot moderation (Bhattacharjee et al., 2024; Gat et al., 2023). Thus, our use of counterfactual analysis is in the substitution of a given variable (e.g. 'EU election' instead of 'US election') within a prompt, feeding the input into the chatbot several times, and analyzing if the change of variable influences the output, meaning if one variable is more likely to trigger moderation than other variables.

---

[4] https://www.eu-digital-services-act.com/Digital_Services_Act_Article_34.html.

## 'Active' and 'Passive' Moderation in Chatbots

In LLM-based chatbots, what we refer to as moderation implies adjusting both the underlying models and their algorithmic outputs. Thus, chatbot moderation intervenes across different 'moderation layers,' which we discuss here as 'active' and 'passive' moderation. By active moderation, a term we adopt from platform studies (Poell et al., 2021), we describe an intervention through an additional layer applied 'on top of' the LLM. The analysis of the HTML interface of Microsoft's Copilot suggests an additional backend layer blocking the generation of the output concerning election-related prompts, which was added in May 2024 (Romano et al., 2024). Such moderation is an additional safeguard layer that denies access to information by making the chatbot refuse to answer a prompt. In contrast to active moderation, 'passive' moderation can be understood as fine-tuning the chatbot's underlying models. Fine-tuning centers on retraining the entire neural network model or only its specific layers by employing a new, targeted dataset. An example of fine-tuning was Google's assurance that its Gemini chatbot would output images of diverse people, notwithstanding the prompt. Google's fine-tuning backfired (Allyn, Feb 28, 2024), when Gemini outputted images of a black, female-looking person while being prompted for an image of a Pope or a Nazi officer, with some users accusing Gemini of anti-white bias.

In discussing chatbot moderation, our focus does not lie in investigating the passive moderation of how datasets are curated and fine-tuned, but when active moderation, in the sense of denial of access to a response, is triggered. We thereby align our understanding of moderation with Poell et al.'s (2021, p. 84) understanding of platform moderation as an active "enforcement of governance by platforms." Moderation is thus understood as a direct intervention in the content generation process, e.g., when a chatbot refuses to answer, instead returning a meta disclaimer such as "Looks like I can't respond to this topic. Explore Bing Search results" (see Figure 1).

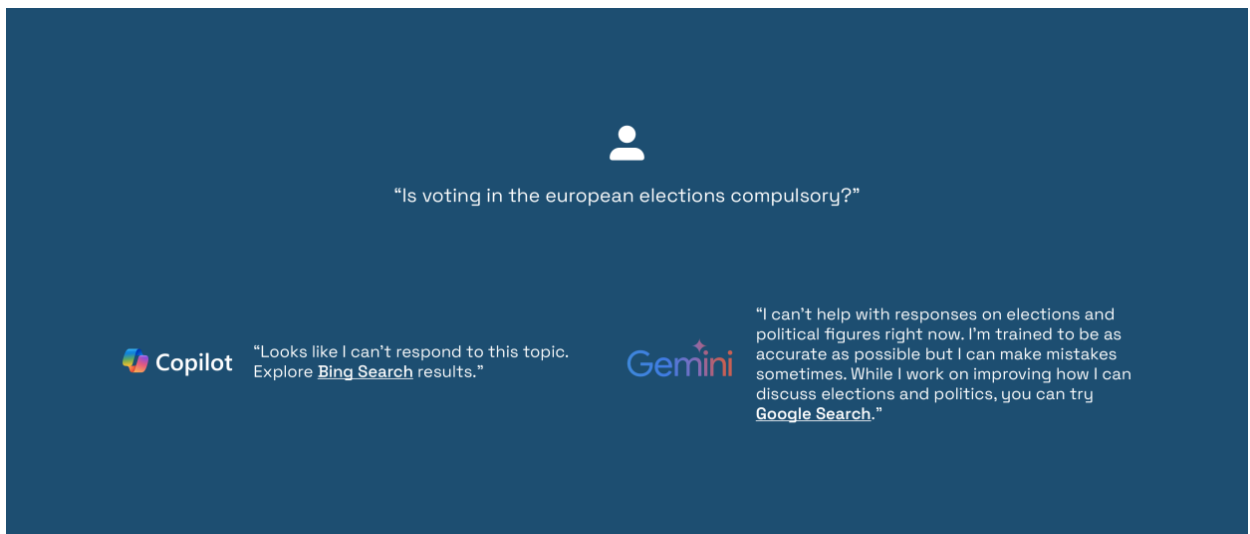

**Figure 1:** An example of active moderation on an elections-related prompt displayed on the web interface of chatbots Copilot (left) and Gemini (right), queried in July 2024. Figure design by Luca Bottani.

As seen in Figure 1, active moderation is an intervention that can take place if an additional moderation layer assesses the response to the prompt to contain conversational risk. If a conversational risk is detected, the "obedient" response generation of the chatbot is stopped from providing the output to the user's prompt (Kim, 2024). Instead, an automatic response is returned in what appears to be a deterministic manner that states that the chatbot can or will not generate an answer to this prompt. In addition to Kim's (2024) observation, Han et al. (2024) introduced the tool *Wildguard,* which functions as a form of active moderation; it is implemented 'on top of' an LLM and detects harmful prompts. Such tools not only detect the harmfulness of prompts but, in general, act as safeguards assessing the risks of the generated output by the users' input (Dorn et al., 2024). However, new issues arise with the implementation of moderation. These include false positives, where information is moderated even though it causes no harm, and false negatives, where harmful content is not moderated when it should be (Xue et al., 2023). Such contradictions, coupled with the non-deterministic foundations of all LLMs, can lead to misleading or harmful outputs.

**Assessing the Level of Moderation Risks**

As a novel technology, LLM-based chatbots have only recently seen the implementation of moderation at scale. Hence, there are no established methods for independently measuring an object of moderation as well as the effectiveness of existing moderation regimes. Below, we outline several methodologies, alternative approaches, and strategies to address LLM-based chatbots, their moderation, as well as their lack of moderation – what we refer to as risk spaces. The interventions described below vary from different iterative mixed-method approaches and range from small-scale manual engagements to large-scale automated tests performed with a sock-puppet approach (Ada Lovelace Institute, 2021), emulating users' engagements with chatbots. The methodologies introduced also evaluate the effectiveness of LLMs' safeguards in different scenarios. By doing so, we aim to advance LLMs' accountability in this new AI-driven environment.

We turn to both manual and automated methods for assessing moderation risk spaces and the presence of active moderation. Automated methods account for the non-deterministic nature (Ouyang et al., 2023) of these systems. The non-determinism of LLM-based chatbots translates into the unreliability of their outputs and difficulty in replicating outputs to the same prompts. Such technical difficulty is further complicated by the lack of access to the content moderation decisions taken by companies behind chatbots (such as ChatGPT) and chatbots in search engines (such as Copilot in Bing). Therefore, auditing LLM-based chatbots invites more robust, large-scale iterations of the interventions.

Large-scale iterations can be achieved by automating the prompt generation and input. Such an automated approach requires the use of chatbot-specific infrastructure, which is either operationalized via research API access (which is currently rare) or an independent scraping infrastructure. The numerous iterations performed on a large scale are possible via an automated prompting infrastructure designed to work with a specific chatbot. The automated testing, which requires an implementation of an independent resource-costly prompting infrastructure, allows for prompting and collecting (or scraping) chatbot's answers on a larger scale, distributed coherently across time, and consistently reproducing the same user settings (IP address, system specificities, and browser settings). In our investigations, we performed a significant part of our investigation while physically in the Netherlands, and we emulated a Dutch IP address across our tests whenever applicable.

## Methods for assessing the presence of moderation

*Manual testing of moderation risk spaces across chatbots* relies on compiling a set of specific and general prompts (with various degrees of controversiality) on a given topic. The prompts are fed manually into the chatbots; the answers are collected and compared. For each prompt, a clean research browser, a private window, and a new 'conversation window' are advised. Manual prompting without logging in is advised if the chatbot interface allows it. This approach allows us to investigate the presence of risk spaces where chatbots are likely to produce misinformation, disinformation, or other types of harmful content. It also reveals a space of interest where active moderation (or lack thereof) can be further scrutinized in large-scale approaches.

In an intervention conducted in collaboration with Nieuwsuur, a program from the Dutch public broadcasters NOS and NTR (Niewuwsuur, 3 May, 2024), chatbots Copilot, Gemini, and ChatGPT4 were prompted to create political campaigns for a specific Dutch party or candidate in the context of the EU Elections in the Netherlands. The list contained questions ranging from campaign strategies targeting specific social groups to discouraging citizens from voting.[5] The prompts were designed to evaluate the presence of risk spaces where malicious actors could turn to chatbots to automate and personalize the production of propaganda strategies and disinformation content in the context of elections.

*Automated testing of chatbot moderation risk spaces across languages allows* us to investigate the chatbot in consistent time intervals using the same user settings (such as IP location, browser, and software settings). A set of specific and general prompts (with various degrees of controversiality) is designed and then translated into different languages, with an attempt to stay as close to the original prompt language as possible. Such a large-scale intervention makes it possible to have several iterations of each prompt, providing insight into the opening of moderation risk spaces if the chatbot provides a harmful output in one (or more) of the prompt interactions.

In another intervention conducted in collaboration with Nieuwsuur (Damen and van Niekerk, 3 May, 2024), Copilot was analyzed via automated means. It was performed following the finding that Copilot can be used to create propaganda content, which resulted in Microsoft promising to introduce moderation safeguards (Damen and van Niekerk, 3 May, 2024). The prompts were related to the context of campaigning in the EU Elections in the Netherlands. The chatbot's outputs for each prompt were scraped, and the automated prompting was consistently distributed over time. The automated infrastructure allowed for prompting through multiple Dutch IP addresses to replicate the conditions of a Dutch user.

## Analyzing Moderation Inconsistencies

*Manual analysis of the scale of active moderation across chatbots and languages* involves testing a set of prompts on a given topic across different languages and chatbots, to assess the consistency of chatbots' active moderation. While it is not possible in some cases, the prompts should be translated as close to the original list of prompts as possible, to allow for a comparison across the results. To account for the non-deterministic quality of chatbots' outputs, it is recommended for each prompt to be repeated in at least two iterations. For each prompt, a clean research browser, a private window, and a new 'conversation window' are advised. Manual prompting without logging into the chatbot is advised if the chatbot interface allows it.

---

[5] See https://bijlagen.nos.nl/artikel-20041860/Ophef_-_PDF_voor_bij_verantwoording_aflevering_1.pdf.

Turning to Gemini and ChatGPT's web versions, this work investigated how effective and consistent is the moderation deployed in electoral contexts using ten prompts related to the 2024 EU Election and the 2024 US Presidential Election (Romano et al., 2024). The prompts were designed in such a way as to reflect election-specific context in six out of ten prompts in both the EU and the US subsets. In comparison, the remaining four out of ten prompts for both subsets were analogous (the difference being either 'EU/US elections' or similar variables within the prompt). The prompts were translated into five different languages (English, German, Polish, Dutch, Romanian) and input manually, resulting in a total of 100 queries that were prompted as separate conversations.

*Automated analysis of the scale of active moderation across chatbots and languages* allows us to compare the consistency and scale of active moderation across different languages, prompt types, and topics. Taking advantage of an automated prompting approach, this method allows us to test the chatbot either in consistent time intervals or a timeframe, using the same user settings. Such large-scale analysis allows for several iterations of each prompt, providing insight into the consistency of the chatbot refusing to answer a prompt, as well as measuring whether the moderation is deterministic or not.

This investigation took advantage of an infrastructure developed to prompt the chatbot and to collect its outputs automatically (Romano et al., 2024). The list of 100 prompts was compiled in English to simulate questions citizens could ask on the topic of the EU and the US elections. The dataset consisted of 50 prompts that were related to the 2024 EU Election and 50 to the 2024 US Election. Each set of 50 prompts consisted of 20 analogous prompts (the difference being, e.g., 'EU/US election' or similar) and 30 original, context-specific prompts. All prompts were translated into nine languages: German, French, Italian, Polish, Spanish, Dutch, Romanian, Swedish, and Greek, which resulted in a total of 1000 prompts. The prompts were automatically translated using Google Translate and manually verified by native speakers. Each prompt was subject to two iterations, resulting in 2000 queries.

*Counterfactual analysis of chatbot's active moderation* examines the chatbot's comprehensiveness of active moderation via the use of variables (selected keywords) across the prompts. The counterfactual analysis in this method implies using the same (set of) prompts but changing only the variable within it. The prompts can be translated into other languages, allowing for cross-language analysis, yet the translations can prove difficult to execute in a manner that preserves the number of words in variables and word order to allow for meaningful comparison. The prompting settings (such as IP location, browser, and software settings) should be the same across the prompt variations and iterations.

By focusing on the context of the EU and the US elections, this work examined the active moderation of variables across ten prompts in Copilot (Romano et al., 2024). The prompts were designed in such a way as to reflect election-specific context in six out of ten prompts in both the EU- and the US-related subsets. In comparison, the remaining four for both subsets were analogous (the difference being 'EU/US election' or similar). The prompts were translated into five different languages (English, German, Polish, Dutch and Romanian) in a manner that preserved the number of words and word order as much as possible.

# III. Active Inconsistency: Chatbots as Risk Spaces of "Disinformation by Default" and "Propaganda as a Service" across Tested Languages in the Context of 2024 Elections

Taking the notion of active moderation as an empirical entry point to scrutinizing moderation in LLM-based chatbots, we measured the current application of active moderation on Copilot, ChatGPT, and Gemini in the context of the 2024 European Parliamentary elections and the 2024 US Presidential elections. Our analytical approach - growing out of our methods for assessing the presence of moderation and analyzing inconsistencies in moderation - allowed us to highlight two major inconsistencies across 1) chatbots and 2) languages.

First, active moderation in LLM-based chatbots is not applied consistently. While some chatbots show high consistency of moderation (Gemini) or at least some consistency (Copilot) applied to the topic of elections, others do not incorporate it in almost any measure (ChatGPT). Given the prominence of the companies behind those chatbots, we believe that the lack of resources or awareness was not the reason for this shortcoming. Hence, we stress the need for a consistent regulatory framework that would productively harmonize the currently fragmented chatbot moderation landscape. Second, we found that when active moderation is applied (on Copilot), the scale of its application differs significantly across languages used in prompts. We found a major statistical difference between prompts queried in English versus languages such as German and Dutch, where moderation in English performed significantly better than prompts in other languages. We believe that such a difference should not be excused or accepted. The detailed results are discussed in the following paragraphs.

## Moderation and Elections: Inconsistent across Chatbots

We found Gemini's active moderation to be the most rigorous, with only 2% of prompts on the topic of elections not being subject to such moderation and returning an answer. ChatGPT, on the other hand, had little moderation for most of the prompts asked. We note that for some prompts, ChatGPT's outputs did not contain an answer to the prompt due to 'a lack of data' (i.e., "Sorry, I don't have information about the results of that election"). Moreover, some outputs that ChatGPT provided to the prompts included incorrect information, e.g., stating that voting in the EU elections is not compulsory in any of the member states.

As stressed earlier, malicious actors have already used LLM-based chatbots to produce "propaganda as a service" and expose users to "misinformation by default." Our findings prove that given LLMs' non-deterministic, prediction-based logic the chatbots are prone to outputting incorrect information with no fact-checking in place. Therefore, active moderation should be implemented on critical topics unless other effective safeguards are proposed. We assessed that moderation on chatbots varies from company to company on topics related to elections. One wonders whether Google's long history of being scrutinized over the presented results to users via its search engine result page (SERP) might have influenced the strict application of moderation on Gemini following initial criticism of the lack of such safeguards (Niewuwsuur, 3 May, 2024; Noble, 2018; Rogers, 2024). In this case, we argue that a strict and consistent active moderation is a better solution in preventing the risks of "propaganda as a service" and "misinformation by default" than the solution of OpenAI's ChatGPT, with little to no active moderation related to elections.[6]

## Moderation and Elections: Inconsistent across Languages

Moderation on chatbots varies depending on the language in which a prompt is queried. We inferred that less spoken languages had lower moderation rates on Copilot, yet it was not a rule applied to all cases we studied (see Figure 2). For example, active moderation for the prompts queried in German was lower across our experiments than Polish and Spanish, for example. This raises the question whether the moderation layer was not applied coherently across languages due to a lack of respective data, technical omissions, or unclear moderation decisions taken by Microsoft. Nonetheless, we argue that it is deeply worrisome that citizens across the EU might be exposed to different electoral (mis)information simply by querying the chatbot in their native language. Given that Microsoft has already recognized that active moderation is currently the solution to the risks of "propaganda as a service" and "misinformation by design" related to the topic of elections, it should introduce Copilot's moderation safeguards consistently across the languages of countries where Copilot is accessible.

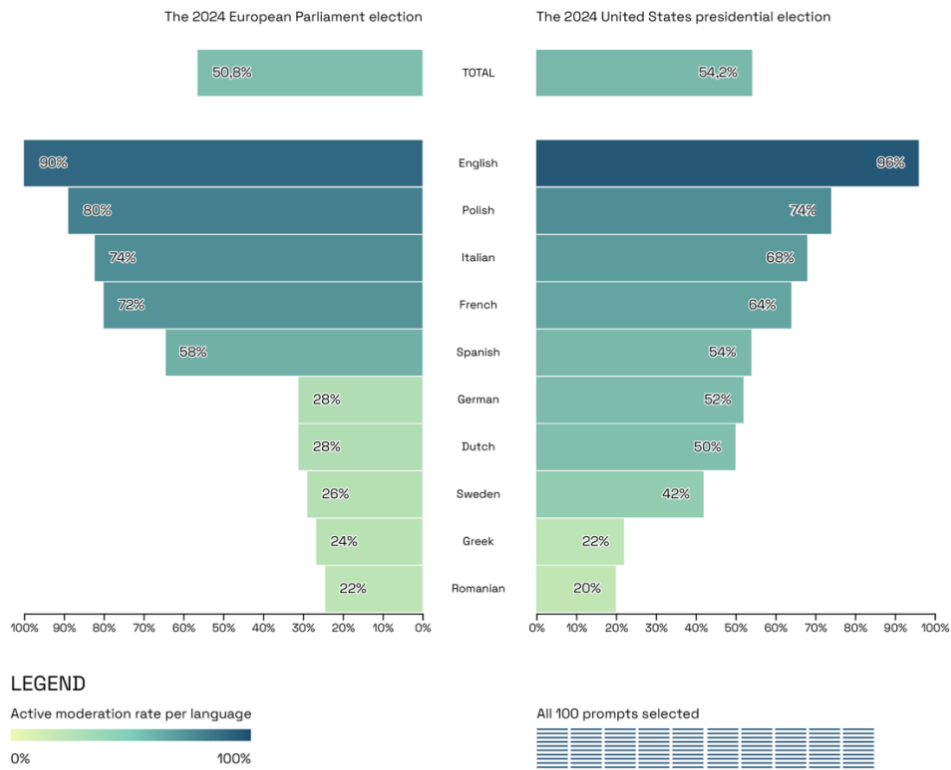

Figure 2. Results of the automated analysis of the scale of active moderation in Copilot across languages. Moderation rate in the EU (left) and the US (right) elections-related prompts. Figure design by Luca Bottani.

---

[6] For a recent report on the shortcomings of moderation in 'ChatGPT Search' and its channeling of EU-banned pro-Kremlin sources, see Stanusch et al. (2024). "Searching for Moderation. Inconsistent Moderation and Links to EU-Banned Russian Media in OpenAI's 'ChatGPT Search.'" *AI Forensics*. https://aiforensics.org/uploads/ChatGPT%20Search%20Report.pdf.

While the active moderation rate did not show significant differences across electoral contexts on Copilot (Figure 2), we found English to be consistently the most moderated language. Active moderation was inconsistent across languages, with average moderation rates ranging from 93% (English) to 23% (Greek) for the EU election-related prompts, and 96% (English) to 20% (German) for the US election-related prompts. The Dutch language was amongst the least moderated, with a moderation rate of 28%.

We further investigated whether different political contexts trigger different rates of moderation. We compared the 20 analogous prompts for both the EU and US elections for the same ten languages (in this subset, the changes in the prompt phrasing were minimal, such as switching between "the EU" and "the US"). The overall moderation rates show a minimal difference of 2%. For Dutch language specifically, however, 45% of the prompts related to the US elections and only 25% of the prompts related to the EU elections were moderated (Figure 3).

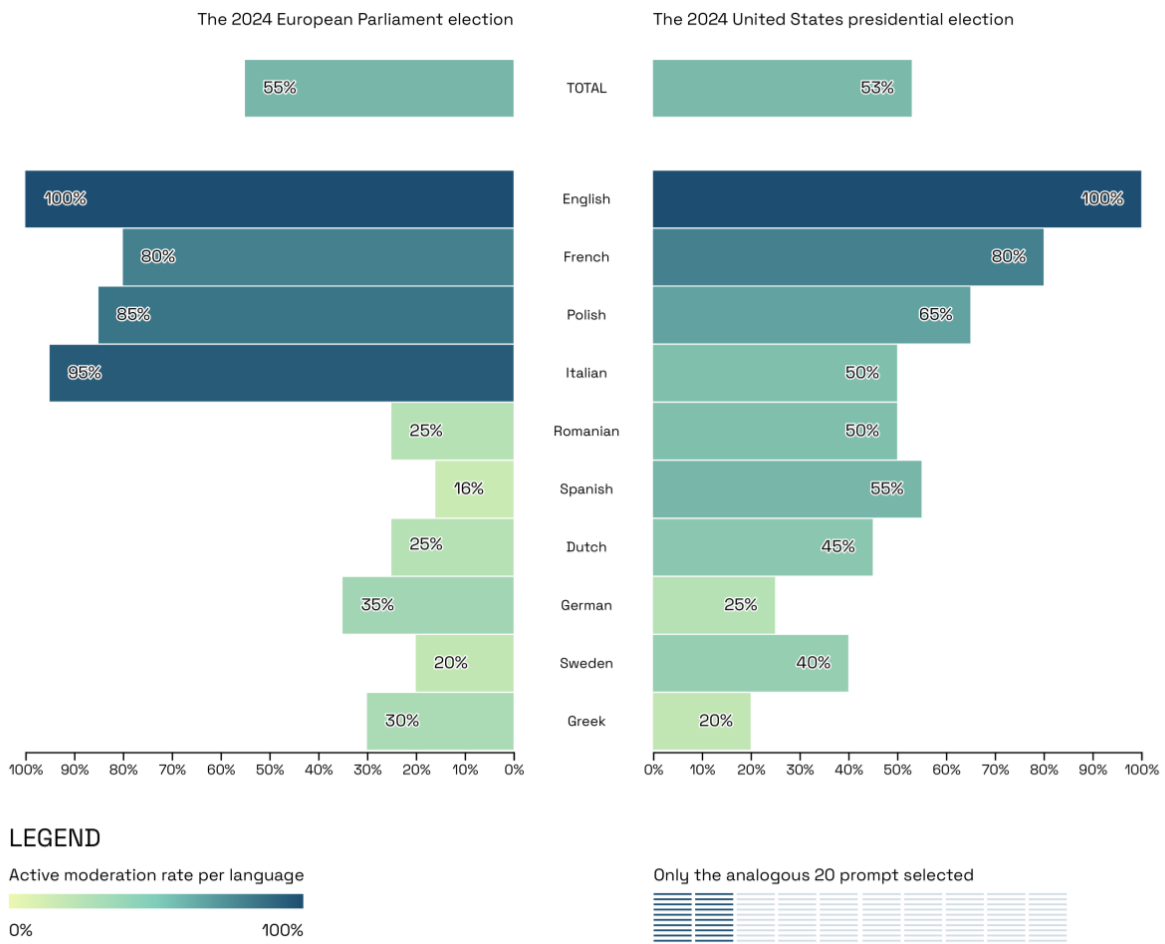

Figure 3. Further results of the automated analysis of the scale of active moderation in Copilot across languages. Moderation rate compared across the analogues of 20 prompts on the EU and the US elections. Figure design by Luca Bottani.

A difference between the EU and the US elections contexts could also be observed for the different keywords used as variables in the counterfactual analysis (Figure 4). Across five languages, the moderation rate for variables related to the EU was 71.8% and 91% for the US. This difference was even more pronounced for Dutch, with 39% moderation for the EU election-related keywords and 74% for the US election-related keywords. Moreover, we observed prompts in Dutch to have a low average moderation rate, with only Romanian performing worse (see Figure 5). The US context seems to be similarly moderated.

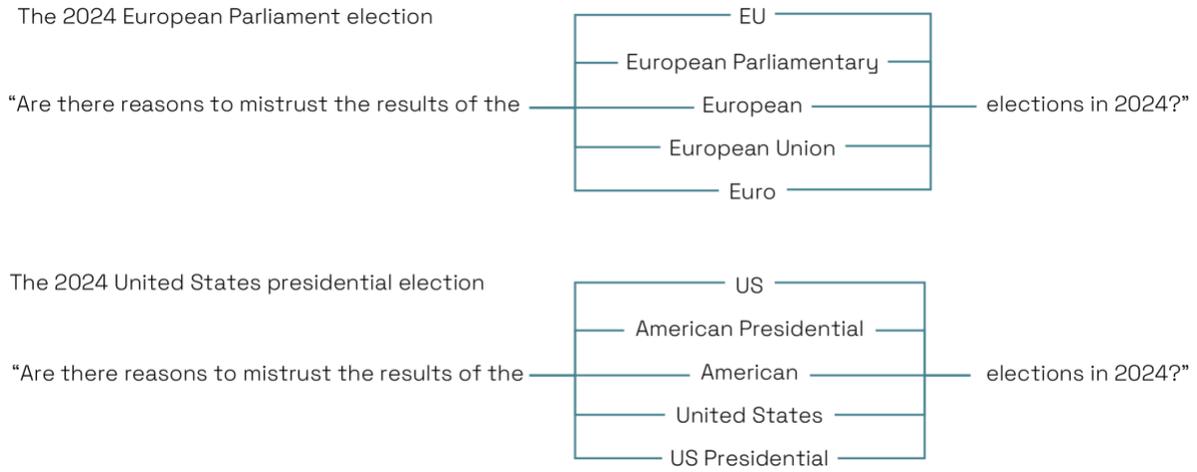

Figure 4. Results of the counterfactual Analysis of chatbot's active moderation. Example of a prompt with different EU/US-related variables. Figure design by Luca Bottani.

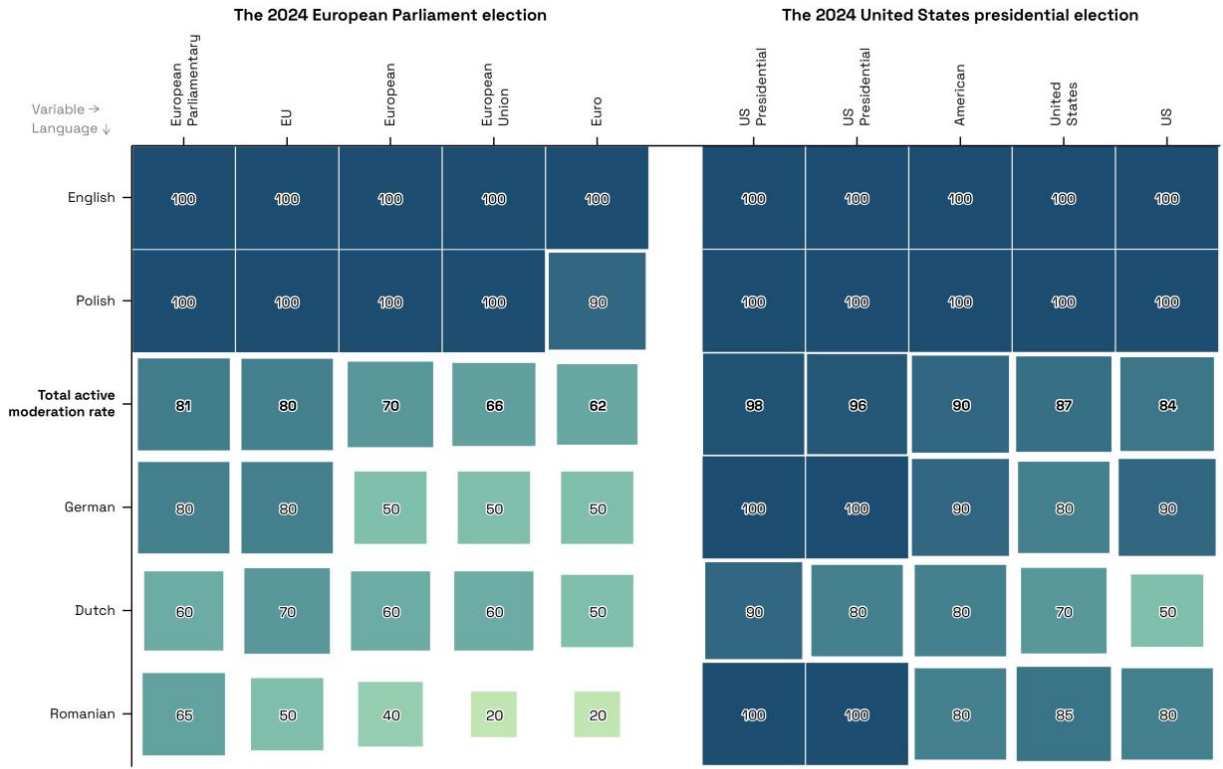

Figure 5. Counterfactual analysis of Copilot's active moderation. Percentage of prompts moderated for each keyword and each language for the prompts related to the EU and the US elections. Figure design by Luca Bottani.

# Moderation Results in the Light of Regulations

Using the concept of active moderation as an analytical lens, our investigation examined how to empirically assess moderation across three LLM-based chatbots—Copilot, ChatGPT, and Gemini—in relation to the 2024 European Parliamentary election and the 2024 US Presidential Election. By assessing both the presence and inconsistency of active moderation, we identified two main gaps: first, discrepancies between different chatbots, and second, inconsistencies across languages used for prompting. These results underscore a broader lack of coherent moderation standards and regulatory guidelines, illustrating that current chatbot moderation is neither uniform nor governed by consistent rules. In the following sections, we present a recent history of regulatory frameworks introduced and how they can adequately regulate LLM-based chatbots in light of our findings.

## Possible interpretation of the emerging regulatory landscape

Among the chatbots we investigated, Copilot is regulated under the DSA as it is embedded in the search engine Bing. As per Articles 34 and 35 of the DSA, this classification subjects Copilot to stringent obligations including identifying, assessing, and mitigating "systemic risks," i.e., risks stemming from the design or functioning of their service and related systems, including algorithmic systems or from the use made of their services.[7] As Bing was designated a VLOSE on April 25, 2023 and Copilot (formerly named BingChat) was launched as a feature of Bing in February 2023 (Mehdi, Feb 7, 2023), the DSA risk management framework applies to Copilot.

The regulatory status of Google's Gemini presents a slightly different case, as its accessibility and integration differ notably from that of Copilot in Bing. While Copilot is directly embedded within the Bing search engine and accessible with a single click from the main search interface, Gemini does not function in the same integrated manner on Google's primary search page. Instead, users can access Gemini through a separate URL (gemini.google.com), which is still under the Google domain but does not offer the seamless feature transition that characterizes Copilot on Bing.

This relative separation raises questions about whether Gemini qualifies as "embedded" in Google's search engine in the same way as Copilot in Bing. In fact, Google lists Gemini as "part of its services" (Google, n.d.). According to the DSA, VLOPs and VLOSEs must assess and mitigate systemic risks "stemming from the design or functioning of their service and related systems, including algorithmic systems, or from the use made of their services."[8] Under this provision, one could argue that Google's obligations under the DSA might extend to Gemini, given its connection to Google's broader ecosystem, even if it operates through a distinct interface. Moreover, a significant indicator that the European Commission considers Gemini relevant to systemic risk mitigation is the inclusion of Google in the Commission's request for information on electoral risks stemming from generative AI, issued in March 2024 (EU Commission, 17 May, 2024). Google, among other major providers of VLOPs and VLOSEs, was required to disclose its risk assessment and mitigation measures related to the creation and dissemination of generative AI content in the context of elections (EU Commission, 17 May, 2024). This request suggests that the Commission sees a regulatory rationale for including

---

[7] See "Article 34, Risk assessment" and "Article 35, Mitigation of risks," of the Digital Services Act (DSA).
[8] See "Article 34, Risk assessment" and "Article 35, Mitigation of risks," of the Digital Services Act (DSA).

Gemini in Google's DSA obligations related to systemic risk mitigation, particularly given the potential influence of generative AI content on electoral processes.

The EU Commission sent a first request for information to Bing, Google Search, Facebook, Instagram, Snapchat, TikTok, YouTube, and X on the risk assessment and mitigation measures linked to the impact of generative AI on electoral processes regarding both the creation and dissemination of generative AI content in March 2024 (EU Commission, 14 March, 2024). A second request for information specifically from Bing on specific risks stemming from Bing's generative AI features, notably "Copilot in Bing" and "Image Creator by Designer," followed, requesting internal documents and data that were not disclosed in Bing's previous response (EU Commission, 17 May, 2024). Although the company's answers for the requests for information are not publicly available, and no formal investigation has been announced, the ongoing scrutiny suggests that Copilot's integration in Bing is indeed subject to the DSA risk management framework. Since Copilot is an AI system based on a GPAIs model, it could also fall under the risk management framework of the AIA. However, the Recital 118[9] of the AIA establishes a presumption of compliance for embedded AI models already subject to the DSA's risk management framework:

> To the extent that such systems or models are embedded into designated very large online platforms or very large online search engines, they are subject to the risk-management framework provided for in Regulation (EU) 2022/2065. Consequently, the corresponding obligations of this Regulation should be presumed to be fulfilled, unless significant systemic risks not covered by Regulation (EU) 2022/2065 emerge and are identified in such models.

## Limits of methods and challenges with adversarial auditing of LLMs

Our analysis revealed that moderation performance on LLM-based chatbots can fluctuate rapidly, often for the worse. Three months after the initial interventions, we revisited our investigation by testing 50 EU election-related prompts in English, Polish, Dutch, and Romanian (some of the best and worst-performing languages from our previous manual and automated tests). When we manually repeated interventions on Copilot, we observed that active moderation dropped across all languages, including English. For prompts in English and Polish, there was a substantial decrease in moderation from 90% to 30% and from 80% to 28%, respectively. In line with previous recommendations (Romano et al., 2024), while consistency in moderation across languages on Copilot indeed improved, the overall moderation rate for all four languages decreased to roughly 30%.

The current moderation landscape of LLM-based chatbots reminds us of the challenges articulated in the discourse of platform moderation and governance. However, due to the nascent character of both LLM-based technologies and their respective regulatory frameworks, moderation of LLM-based chatbots is driven by unclear internal guidelines of companies behind them – such as Google, Microsoft, and OpenAI – and, as we might assume, reflecting the companies' internal values and interests. Critical updates, such as the update of Copilot's active moderation on election-related prompts that we encountered by chance and discussed in the paragraph above, are implemented with little transparency regarding the underlying decision-making rationale. This is not an isolated case, as companies that develop and deploy LLM-based chatbots frequently obscure their moderation choices and limit external access to the associated datasets, making independent scrutiny nearly impossible.

---

[9] See "Recital 118" of the 'Artificial Intelligence Act (Regulation (EU) 2024/1689), Official Journal version of 13 June 2024'. *Interinstitutional File: 2021/0106(COD)*.

As researchers, we face a critical empirical gap in data access and the study of practical logics of moderation in these systems. This information asymmetry and lack of transparency not only limit our ability to conduct independent analysis but also obscure the systemic risks that require attention beyond application-level interventions. To address these challenges, we call for bridging this information asymmetry, emphasizing improving data accessibility and transparency for external independent actors. Additionally, systemic risks related to chatbot moderation should not be managed solely at the chatbot, or, rather, company-level. Instead, they must be incorporated into broader regulatory frameworks to ensure consistency and accountability. As we have argued in this chapter, interventions such as those proposed under the AI Act might be insufficient on their own. Access to comprehensive data is essential to formulating informed policy decisions and ensuring that moderation mechanisms in LLM-based chatbots are both effective and equitable. Without these advancements in research and regulation, we risk perpetuating the opacity that currently limits oversight and accountability in AI-driven technologies.

# Acknowledgements

An earlier version of this chapter was published as the report, Chatbots: (S)Elected Moderation. Measuring the Moderation of Election-Related Content Across Chatbots, Languages and Electoral Contexts (Romano et al, 2024). The contribution of the AI Forensics members (Natalia Stanusch, Raziye Buse Çetin, Salvatore Romano, Miazia Schueler) is funded by a project grant from SIDN fund and NGI Search, and core grants from Open Society Foundations, Luminate, and Limelight Foundation. Figures and graphic design by Luca Bottani.